\documentclass[prb,12pt,endfloats]{revtex4-1}
\usepackage{color}
\begin{document}
%\draft
\title{Vibrational Behavior of Metal Nanowires under Tensile Stress}
\author{Yasemin \c{S}eng\"un$^1$ and Sondan Durukano\u{g}lu$^{2,3}$}
\affiliation{$^1$Department of Physics, Istanbul Technical University, Maslak, 34469 Istanbul, Turkey\\
$^2$Faculty of Engineering and Natural Sciences, Sabanci University, Orhanli, Tuzla 34950 Istanbul, Turkey\\
$^3$Feza G\"ursey Research Institute, T\"UB\.ITAK-Bo\u{g}azi\c{c}i ~University,  \c{C}engelk\"oy, 34684 Istanbul, Turkey }
\date{\today}

\begin{abstract}
We have investigated the vibrational density of states (VDOS) of a thin Cu nanowire with $<$100$>$ axial orientation and considered the effect of axial strain.  The VDOS are calculated using a real space Green's function approach with the force constant matrices extracted from interaction potential based on the embedded atom method. Results for the vibrational density of states of a strain-free nanowire show  quite distinctive characteristics  compared to that of a bulk atom, the most striking feature of which is the existence of high frequency modes above the top of the bulk spectrum.  We further predict that the low frequency characteristics of the VDOS reveal the quasi-1 dimensional (Q1D) behavior only when the wire is extremely thin.  Through decomposition of VDOS into local atoms we also exhibit that while the anomalous increase in low frequency density of states is mainly due to the corner atoms, the enhancement in high frequency  modes is primarily moderated by core atoms.  We, additionally, find that while the high frequency band above the top of the bulk phonon shifts to higher frequencies, the characteristics at low frequencies remains almost the same upon stretching the nanowire along the axial direction.

\end{abstract}
%PACS\#
\maketitle
\newpage

Nanostructured materials are excellent examples for investigating finite size effects on the physical properties of a system of interest.  One such effect is the novel  characteristics observed in the vibrational behavior of these materials in comparison with the corresponding  bulk and surface systems.   The driving factors for such modifications in vibrational density of states (VDOS) may be  varying characteristics in force fields experienced  by individual atoms, and thus distinct vibrational contribution of every atom or the dimensional confinement effects on the phonons of the system.  Key questions concerning the thermal conductivity, electron transport,  and thermal stability of nanostructured materials are clearly related to  modifications in VDOS induced by the finite size effects that are very often manifested in novel characteristics of the system like an increase of low frequency modes, higher vibrational amplitudes of the  atoms  and thus excess vibrational entropy.   For example,  the voltage-dependent conduction of metallic, monoatomic wires is found to be associated to inelastic scattering of electrons with phonons\cite{Agrait}.  Recently, it was suggested that  formation of CeH$_{2.84}$ nanoplates by fracture is due to the excess vibrational entropy introduced by the instantaneous phonon confinement as the crack propagates\cite{Manley}.  Furthermore, both experimental and theoretical studies on metallic nanocrystals have revealed an enhancement of the phonon density of states (DOS) at low and high frequencies\cite{PhonFe,PhonNi3Fe,PhonNi,PhonNiCu,Kara98}.  Thus studies of vibrational dynamics are essential to have a complete understanding of varying characteristics and thermodynamic stability of nanocrystals.

For nanocrystalline materials$ -$ the inherent hosts of grain boundaries, the enhancement in the low and high frequency modes is attributed  to the grain boundaries where the force fields experienced by the local atoms are expected to be quite different compared to those in regular or denser domains of the crystal\cite{PhonNiCu}. Even more interestingly,  molecular dynamic simulations on the helical, ultrathin gold nanowires that are under stretching reveal that while the low frequency modes in phonon density of states shift to lower frequencies upon stretching, the high frequency modes display  almost no significant change\cite{PhonHelixAu}.  Clearly, the arrangements of  atoms within the nanocrystal, and thus the local force fields have a crucial  role in determining the varying characteristics of VDOS.  In this Report, through reliable calculations we have examined the VDOS of a rectangular, thin Cu nanowire and have found that the finite size effects on the vibrational characteristics are significant  and  these varying behaviors can be traced to partial or local contribution of the atoms of the cross-sectional plane with different atomic environment.

Here, we are interested in a rectangular Cu nanowire with $<$100$>$ axial orientation that can be fabricated by cutting the
(100) slab through the (100) walls of a square.  As indicated in Fig. 1, in this particular type of nanowire the cross-sectional area
is a square and specified by the number of atoms along the diagonals, namely by 5$\times$5.    In our calculations the interaction between the atoms is defined using a vastly tested, semi-empirical interaction potential based on the embedded atom method (EAM)\cite{EAM}.  These potentials have so far proven to be reliable for examining the energetics, structure, and dynamics of low-coordinated surfaces and nano-structured materials\cite{Duru1,kim,Duru2}.  We have taken the $x$ and $y$ axis to lie in the cross-sectional plane and $z$ being along the axial direction of the wire.  In the simulations, periodic boundary condition is applied along the axial direction to simulate an infinitely long nanowire, while no such constraint is imposed along the directions parallel to the cross-sectional plane.

To accurately determine the vibrational density of states, we use a real-space Green's function
technique which is particularly suited for systems with defects, disorder, and reduced symmetry
as it does not require the system to be periodic\cite{RSGF}. The only prerequisite to this method is that
the interaction potential between the atoms in the system be of finite range.  The normalized vibrational
density of states associated with locality $l$ is then given by

\begin{equation}
n_{l}(\omega)=2\omega g_{l}(\omega^2)
\end{equation}
\noindent
where the function $g_{l}(\omega^2)$ satisfies the equation
\begin{equation}
g_{l}{(\omega^2)} = - {1 \over 3N_l\pi} \; \lim_{\epsilon\rightarrow0}
\left\{ Im[{\rm Tr}(G_{ll}(\omega^2+i\epsilon))] \right\}.
\end{equation}
\noindent
In the above equation $G_{ll}$ is the Green's function
matrix corresponding to locality $l$ and $N_l$ is the number of
 atoms in this locality.  Then sum over all atoms in the system gives the total VDOS corresponding to the system.  Full details for calculating the Green's
function corresponding to the local region of interest are
described in Ref.\ 13.

 Very briefly, the atoms in our model system are initially arranged in their perfect lattice
positions of the truncated nanowire and allowed to interact via EAM potentials.  Next, the standard conjugate gradient method is used to
fully minimize the total energy of the system for eliminating non-zero initial stress.  Once the equilibrated structure
is attained, a series of strain ranging from 0\% to 5\%, with an increment of $0.5\%$, is applied.
The force constant matrix corresponding to each strained and subsequently equilibrated structure is then constructed using the analytical
expressions for the partial second derivatives of EAM potentials.  We have finally calculated the local vibrational
density of states corresponding to any local atom of interest in each strained nanowire.

In Fig. 2(a), we have plotted  total VDOS of the $5\times 5$ type Cu nanowire, together with that of a bulk atom.  As seen in the figure,
there are two distinctive characteristics  compared to DOS of a bulk atom: a strong shift towards low frequency modes which is expected and existence of high frequency modes above the top of the bulk phonon spectrum.  Indeed, presence of such high frequency modes for nanostructured materials has been confirmed both experimentally and theoretically\cite{PhonFe,PhonNi3Fe,PhonNi,PhonNiCu,Kara98}.  However, in the inelastic neutron scattering experiment on Ni nanocrytalline samples\cite{PhonNi} and resonant inelastic nuclear $\gamma $-ray scattering measurements\cite{PhonFe} on nanocrystalline Fe, these high frequency modes are attributed to hydrogen and oxygen contamination, respectively.  In molecular dynamic simulations on nanocrystalline Cu and Ni samples Derlet and co-workers have  identified these modes as an anharmonic contribution coming from the finite temperature\cite{PhonNiCu}.  On the contrary, from our results we conclude that  existence of high frequency modes  may be an inherent characteristic of low dimensional systems and may not be due to contamination or temperature driven anharmonic effects as we carried out the calculations for ground state configuration (T=0K) of a pure Cu nanowire.  However, we recognize that  with temperature anharmonic effects may become important and may even induce higher frequency modes.  Also,  the low frequency behavior  of the VDOS of the wire is found to be linear with frequency instead of conventional quadratic dependence of the bulk, affirming the reduced dimensionality of the system.  On the other hand, for nanowires of that thin one might treat the sytem as a quasi-1 dimensional (Q1D) wire and expect to observe the Q1D-vibrational characteristic of frequency-independent VDOS at low frequencies.  Nevertheless, even the local vibrational density of states corresponding to the center atom on 5$\times$5, a chain of which is practically expected to form a Q1D system, does not display such a behavior (see Fig. 2(b)). To further address this issue, we performed similar calculations on two geometrically different nanowires with smaller cross-sectional area: one is representative of square cross-sectional area with 3 atoms along the diagonals, namely 3$\times$3 and the other one is an example of hexagonal cross-sectional area with $<$111$>$ axial orientation.  In Fig. 3 we have plotted the local vibrational density of states of a center atom on both types of nanowires together with the top views of the associated cross-sectional areas. Although the LDOS for 5$\times$5 type nanowire does not exhibit the Q1D vibrational characteristic, for 3$\times$3 and H3 type nanowires the LDOS of CA clearly shows the expected low frequency behavior$-$frequency independent VDOS at low frequencies (see Fig. 3).

 In order to identify the major contributors to the predicted enhancement of the low and high frequency modes of the total VDOS, we have calculated  local density of states of each atom on the cross-sectional plane of the wire and presented them in Fig. 2(b).   As it is clear from the figure, the primary contributors to the high frequency tail of the VDOS are the center-atom (CA) and diagonal-atom (DA), whereas  the strong enhancement of the low frequency modes is mainly due to the corner atom (KA).   As pointed out in our previous study\cite{Duru2}, broken symmetries very often manifest themselves  in novel characteristics like localized modes around the edges or surfaces.  Indeed, in a theoretical study, Nishiguchi {\em et al.} have investigated the acoustic phonon modes for square and rectangular quantum wires of GaAs using the xyz-algorithm and confirmed the presence of acoustic modes localized at the corners of the wire\cite{LocalModesGaAs}.  The fact that the enhanced low frequency modes are mostly participated by the corner atoms in our model system seems to be echoing existence of edge localized modes.  However, detailed calculations for obtaining phonon dispersion curves together with the displacement vectors, which is beyond the scope of this work, are needed to confirm existence of such localized modes on these particular Cu nanowires. Furthermore, we previously calculated  LDOS of a center atom (CA) of a strain-free, $5\times 5$ type Cu nanowire along the $x-$  and $z-$direction, and concluded  that  the high frequency modes above the bulk band are  due to mostly the
 vibrations along the directions parallel to the cross-sectional plane\cite{PhysicaA}.   The characteristics of these high frequency modes might be similar to the breathing modes$-$ the lowest-order pure-compressional modes $-$ observed in nanotubes as the modes are  within the longitudinal part of the phonon spectrum with the vibrations along the radial direction.

 The predicted characteristics in LDOS  can be well understood in the context of changes in the
 bond-lengths
between the neighboring atoms and in the context of coordination number.  From the surface energy point of view,  the ground state atomic configuration of the nanowires with high surface to volume ratio is governed by the tendency of surfaces to reduce their surface energies which, in turn, leads to unique relaxation patterns for the atoms  and thus varying characteristics in bond lengths\cite{Duru2}. For the 5$\times$5 type nanowire, the bond-length between the CA, which resembles a bulk atom in terms of coordination number (12 compared to that of a corner atom$-$5),  and its first nearest neighbor (DA) along the diagonal of the cross-sectional plane is shortened by $2\%$ whereas that of a DA and a KA is shortened by $5.5\%$ compared to the bond-length in the bulk.     Generally speaking, the smaller the separation between the atoms, the stronger the force of interactions and vice versa.  With a decrease in its bond lengths with neighboring atoms, the CA  and the DA would experience stronger force fields and hence participate in high frequency modes. Furthermore, in contrast to atoms in the bulk and those on flat surfaces, atoms on nanowires have varying local atomic environments, due to additional lack of neighbors.  For example, the lattice sites at the corner of the wire are the least-coordinated ones with a coordination of 5, compared to the other atoms in the cross-sectional plane; SA has a coordination of 8, DA and CA have a coordination of 12 just like a bulk atom in an $fcc$ crystal.  Thus the corner atoms on the wire are likely to experience the least force fields coming from their neighboring atoms.  As a result, the vibrational frequencies of a KA are expected to be shifted towards the lower frequencies as compared to the DOS of higher coordinated atoms such as bulk, center and diagonal atoms.  Indeed, the marked enhancement in the low frequency modes of the respective LDOS is a clear reflection of the reduced force fields of these low coordinated atoms (see Fig. 2(b)).  On the other hand, the high frequency modes observed in the tail of the LDOS of KA are presumably due to the atomic vibrations of corner atoms that are coupled with the high frequency modes of diagonal atoms in that particular part of the phonon spectrum.

Finally, we let the nanowire to be exposed to an incremental axial strain with an increment of 0.5$\%$  and then calculate the total VDOS corresponding to each strained wire.  The respective VDOS together with  the DOS of a bulk atom  are plotted in Fig. 4.  In contrast to what has been reported for the helical,  ultrathin Au nanowires, the characteristics of low frequency modes in total VDOS remains almost the same, whereas the high frequency modes, that are above the top of the bulk phonon spectrum, shift to even higher frequencies, upon stretching the nanowire.   We believe that this substantially different vibrational response of a rectangular Cu nanowire from that of helical Au nanowire may lie in its structural difference since the atomic arrangements within the cross-sectional plane of a rectangular nanowire comprise compact flat planes of the crystal whereas that of a helical nanowire lacks the spatial symmetry.

For nanowires, vibrational response of each atom to the applied strain is expected to be considerably different for symmetry reasons.  In Fig. 5, we have plotted the local VDOS of a CA and a KA for a comparison.  As seen in the figure, with increasing tensile strain along the axial direction the KA starts participating in modes with frequencies above the top of the bulk band, whereas the behavior in the low frequency part of the spectrum stays the same with the leading contribution coming from the KA.    We also find that the shift towards the higher frequency modes in total VDOS is dictated mainly by CA.

 In conclusion, we have explored  the vibrational properties of  a rectangular Cu nanowire and examined the effect of axial strain on the vibrational behavior of the wire.   Taking the specific example of a 5$\times$5 type and pure Cu nanowire at its 0K ground state configuration, we show that the existence of high frequency modes above the top of the bulk phonon spectrum is a reflection of the reduced dimensionality of the system rather than being an end effect of contamination or temperature.  Through  the projection of the total VDOS on local atoms of the wire, we identified the leading contributors  to the enhancement of the modes at low and high frequencies:  while the anomalous low frequency modes are primarily moderated by the lower coordinated atoms (corner atoms), the aberrant high frequency modes are dominated largely by the higher coordinated atoms (center atoms).  In contrast to the case of helical nanowires, the existing aberrant high frequency modes shift to higher frequencies upon stretching the nanowire.  However, the vibrational behavior at low frequencies remains almost the same with increasing axial strain.  In addition, our preliminary calculations indicate that the Q1D-vibrational characteristic of the frequency-independent VDOS may not be observed unless the wire is extremely thin.

This work was supported by the Scientific and Technological Research Council of Turkey-TUBITAK under Grant No. 109T105. Computations were carried out through the National Center for High Performance Computing, located at Istanbul Technical University, under Grant No. 20132007.

\begin{figure}
\caption{Cross-sectional (on the left) and perspective view (on the right) of the $<$100$>$ axially oriented nanowire
with $5\times 5$ number of atoms along the diagonals.
 The darker and lighter yellow spheres show the atoms in A and B type stacking of Cu(100) crystal.  Here KA, DA,  CA and SA, respectively,  stand for the corner, diagonal, center and side atom of the cross-sectional plane.}
\label{Fig1}
\end{figure}

\begin{figure}
\caption{(a) The total vibrational density of states of a 5$\times$5 type Cu nanowire (dotted line).  The solid line corresponds to DOS of a Cu bulk atom.  (b) The local VDOS for a corner atom$-$KA (darker dashed-dotted line), diagonal atom$-$DA (dotted line), center atom$-$CA (dashed line), SA$-$side atom (lighter dashed$-$dotted line), and bulk atom (solid line). }
\label{Fig2}
\end{figure}

\begin{figure}
\caption{The local vibrational density of states of a center atom on 3$\times$3 and H3 type Cu nanowire.  On the left are the top views of the cross-sectional area of both types.  Here 3$\times$3 is representative of $<$100$>$ axially oriented wire whereas H3 is an example of $<$111$>$ axially oriented wire.  In H3, H stands for hexagonal shape of the cross-sectional area and 3 represents the number of atoms along the diagonals.}
\label{Fig3}
\end{figure}

\begin{figure}
\caption{The total vibrational density of states of a strained nanowire (dotted line), together with DOS of a bulk atom. }
\label{Fig4}
\end{figure}

\begin{figure}
\caption{The local vibrational density of states for  (a) a  center atom$-$CA and  (b)  a corner atom$-$KA.  The nanowire is exposed to a strain, ranging from 0\% and 5\% and the corresponding local VDOS are plotted. The solid line is the DOS of a bulk atom.}
\label{Fig5}
\end{figure}

\end{document}